\documentclass[%
 reprint,
 superscriptaddress,
 amsmath,amssymb,
 aps,
showkeys,
]{revtex4-2} 

\usepackage{graphicx}
\usepackage{dcolumn}
\usepackage{bm}
\usepackage{xcolor}

\usepackage{lineno}

\newcommand{\microresonator}{MR}
\newcommand{\microresonators}{MRs}
\newcommand{\clockwise}{CW}
\newcommand{\counterclockwise}{CCW}
\newcommand{\qualityfactor}{Q}
\newcommand{\qualityfactors}{Qs}
\newcommand{\Figure}{Fig.}
\newcommand{\Figures}{Figs.}
\newcommand{\equationM}{Eq.}
\newcommand{\equationsM}{Eqs.}
\newcommand{\CavityRingDown}{CRD}

\usepackage[a4paper, total={7in, 10.3in}]{geometry}

\begin{document}


\title{Interferometric cavity ring-down technique for ultra-high Q-factor microresonators}

\author{Stefano Biasi}
    \email{Corresponding author: stefano.biasi@unitn.it}
    \affiliation{Nanoscience Laboratory, Department of Physics, University of Trento, Via Sommarive 14, Povo - Trento, 38123, Italy}
\author{Riccardo Franchi}%
    \affiliation{Nanoscience Laboratory, Department of Physics, University of Trento, Via Sommarive 14, Povo - Trento, 38123, Italy}
\author{Lorenzo Pavesi}
    \affiliation{Nanoscience Laboratory, Department of Physics, University of Trento, Via Sommarive 14, Povo - Trento, 38123, Italy}

\date{\today}

\begin{abstract}
Microresonators (\microresonators{}) are key components in integrated optics. As a result, the estimation of their energy storage capacity as measured by the quality factor (\qualityfactor{}) is crucial. However, in \microresonator{} with high/ultra-high  \qualityfactor{}, the surface-wall roughness dominates the intrinsic \qualityfactor{} and generates a coupling between counter-propagating modes. This splits the usual sharp single resonance and makes difficult the use of classical methods to assess  \qualityfactor{}. Here, we theoretically show  that an interferometric excitation can be exploited in a Cavity Ring-Down (\CavityRingDown{}) method to measure the ultimate \qualityfactor{} of a \microresonator{}. In fact, under suitable conditions, the resonant doublet merges into a single Lorentzian and the time dynamics of the \microresonator{} assumes the usual behavior of a single-mode resonator unaffected by backscattering. This allows obtaining a typical exponential decay in the charging and discharging time of the \microresonator{}, and thus, estimating its ultimate \qualityfactor{} by measuring the photon lifetime.
\end{abstract}

\keywords{Optical resonator, Cavity Ring Down, Integrated optics, Non-Hermitian physics, Two-level systems}


\maketitle

High/ultra-high quality \microresonators{} find a wide range of applications spanning from fundamental physics \cite{CombScience2018_Kippenberg, Entaglement2020} to narrow line-width laser sources \cite{LaserSource2019_Gundavarapu}. Advances in fabrication processes have led to the realization of increasingly high-performance \microresonators{} \cite{BookSilicoPhotonicDesign}. Over the years, it has been shown ring \microresonators{} with intrinsic \qualityfactor{} up to nearly 260 million by exploiting fully compatible wafer-scale fabrication \cite{260MilionQ_2021Jin}. Furthermore, a record value of 422 million intrinsic \qualityfactor{} has been achieved in a recent work \cite{NatureRing2021_Puckett} by using a silicon nitride ring \microresonator{}. The \qualityfactor{} is affected by intrinsic material properties, by geometrical factors as well as by the fabrication process \cite{Gorodetsky:96}. This last causes surface-wall roughness that induces backscattering of light and, therefore, a simultaneous excitation of clockwise (\clockwise{}) and counterclockwise (\counterclockwise{}) modes \cite{NatureRing2021_Puckett}. As a result, the system does not exhibit a Lorentzian-shaped spectral response but a resonant doublet \cite{Non-Hermitian_Biasi, BackWill}. In this case, the usual estimate of \qualityfactor{} by the measure of the Full-Width at Half Maximum (FWHM) is inaccurate. In addition, with a high/ultra-high \qualityfactor{} the measurement of the spectral line-width requires sophisticated high spectral resolution techniques. Luckily, when \qualityfactor{} increases also the photon lifetime $\tau_{ph}$ increases which makes easy the use of the \CavityRingDown{} techniques \cite{CavityRing_Berden}. Unfortunately, the presence of the backscattering distorces the exponential decay by introducing oscillations on the intensity due to the energy exchange between the coupled \clockwise{} and \counterclockwise{} modes.  

In this work, we show theoretically how the interferometric excitation\cite{Interferometric_BiaFran} can be used to characterize \microresonators{} with ultra-high \qualityfactor{} in the presence of a resonant doublet due to backscattering. In fact, under specific excitation conditions, the doublet merges into a single Lorentzian. This allows using the \CavityRingDown{} technique. Therefore, we compute the time response of a \microresonator{}/bus waveguide system when both the \clockwise{} and \counterclockwise{} modes are simultaneously excited with a rectangular pulse of light. In this case and under appropriate conditions, the time dynamics reduces to that of a single-mode propagating into a \microresonator{} unaffected by backscattering. Consequently, the ultimate \qualityfactor{} of the structure can be estimated from the \microresonator{} charging and discharging times. With ``ultimate'' we mean the \qualityfactor{} in the absence of backscattering \cite{Gorodetsky:96}.

Let us consider a \microresonator{} coupled to a bus waveguide. The surface-wall roughness induces a coupling between the electric field amplitudes of the \clockwise{} ($\alpha_{\rm CW}$) and \counterclockwise{} ($\alpha_{\rm CCW}$) propagating modes. The coupling is modeled by the $\beta_{12}$ and $\beta_{21}$ coefficients within a temporal coupled mode theory \cite{TCMT_Deriv}. Therefore, the dynamic behavior of the field amplitudes is given by \cite{Interferometric_BiaFran}: 
\begin{equation}\label{eq:TCMT}
    \begin{split}
    	i \frac{d}{dt} \!\!   	\begin{pmatrix}
    		\alpha_{\rm CCW} \\
    		\alpha_{\rm CW}
    	\end{pmatrix}
    	\! = \! 
    	&\begin{pmatrix}
    		\omega_0-i(\gamma+\Gamma) & -i\beta_{12} \\
    		-i\beta_{21} & \omega_0-i(\gamma+\Gamma)
    	\end{pmatrix}\!\!
    	\begin{pmatrix}
    		\alpha_{\rm CCW} \\
    		\alpha_{\rm CW}
    	\end{pmatrix}\\
    	&- \sqrt{2\Gamma}
    	\begin{pmatrix}
    		E_{\rm in, L} \\
    		E_{\rm in, R}
    	\end{pmatrix}\,,
    \end{split}
\end{equation}
where $\omega_0$ is the angular resonant frequency of the \microresonator{} while $\Gamma$ and $\gamma$ are the extrinsic and intrinsic damping rates, respectively. $E_{\rm in,R/L}$ are the field amplitudes input in the bus waveguide from the right/left side. The field amplitudes exiting the bus waveguide from the right ($E_{\rm out,R}$) and left ($E_{\rm out,L}$) sides are:
\begin{equation}\label{eq:Eout}
	\begin{pmatrix}
		E_{\rm out, R} \\
		E_{\rm out, L}
	\end{pmatrix}
	 = 
	\begin{pmatrix}
		E_{\rm in, L} \\
		E_{\rm in, R}
	\end{pmatrix}
	+i\sqrt{2\Gamma}
	\begin{pmatrix}
		\alpha_{\rm CCW} \\
		\alpha_{\rm CW}
	\end{pmatrix}\,.
\end{equation}

To compute the time response of the \microresonator{}  to an interferometric excitation we solve the system of \equationsM{} \eqref{eq:TCMT} by using the Green's functions $G_{\rm CW}[t]$ and $G_{\rm CCW}[t]$ . They are defined as the solution of a differential equation when its forcing equals to a Dirac delta function ($\delta[t]$) \cite{Green_Duffy}. Thus, substituting a coherent excitation from both sides ($E_{\rm in,L}[t] = \epsilon _{\text{in,L}} \delta[t-t’]$ and $E_{\rm in,R}[t] = \epsilon _{\text{in,R}} \delta[t-t’] e^{i \phi}$, where $\phi$ is the phase difference between the input fields and $\epsilon_{\rm in,L/R}$ are the input field amplitudes) and taking the Fourier transform of \equationM{} \eqref{eq:TCMT}, one obtains the expression of the spectral Green functions $G_{\rm CW}[\omega]$ and $G_{\rm CCW}[\omega]$\cite{TimeResonator_Biasi}. Then, substituting these solutions in \eqref{eq:Eout}, one determines the spectral Green function of the output fields. In the following, without loss of generality, we focus our attention on the right-side transmitted field: 
\begin{equation}
\begin{array}{r@{\ }c@{\ }l}
G_{\rm out,R}[\omega] &=& \frac{1}{\sqrt{2 \pi }}\epsilon _{\rm in,L}e^{i \omega t'}\Big\{1-\frac{2 \Gamma (-i \Delta \omega + \gamma +\Gamma)}{(-i \Delta \omega + \gamma +\Gamma)^2-\beta
   _{12} \beta _{21}}\Big\} \\
   &+& \frac{1}{\sqrt{2 \pi }} \epsilon _{\rm in,R} e^{i \omega t'} \Big\{\frac{2
   \Gamma \beta _{12}}{(-i \Delta \omega + \gamma +\Gamma)^2-\beta _{12} \beta _{21}}\Big\} e^{i \phi },
\end{array}
\label{eq:SpectralGreenFunction}
\end{equation}
where $\Delta \omega = \omega - \omega_0$ is the detuning (input minus resonant angular frequency). The inverse Fourier transform of the previous equation leads to the Green function as:
\begin{equation}
\begin{array}{r@{\ }c@{\ }l}
     G_{\rm out,R}[t-t']&=&\epsilon _{\rm in,L} \Big\{\delta[t-t'] - \Gamma \Theta[t-t']\\ &\cdot&\left(e^{-i(t-t')\lambda_{1}} + e^{-i(t-t')\lambda_{2}}\right)\Big\}\\
     &+& \sqrt{\frac{\beta_{12}}{\beta_{21}}} \epsilon_{\rm in,R} \Gamma \Theta[t-t']\\
     &\cdot& \Big\{e^{-i (t-t')\lambda_{2}}-e^{-i(t-t')\lambda_{1}} \Big\} e^{i \phi},
\end{array}
\label{eq:TimeGreenFunction}
\end{equation}
where 
$\lambda_{1,2}= \omega_0 - i\left(\gamma +\Gamma \right) \pm \sqrt{- \beta_{12}\beta_{21}}$ are the eigenvalues of \equationM{}\eqref{eq:TCMT} and $\Theta[t]$ is the Heaviside function. Finally, the convolution of $G_{\rm out,R}[t]$ with the exciting function, $\xi_{\rm in}[t]$, yields to $E_{\rm out,R}[t]$ \cite{TimeResonator_Biasi}.

\begin{figure}[t!]
	\centering
	\includegraphics[scale=1]{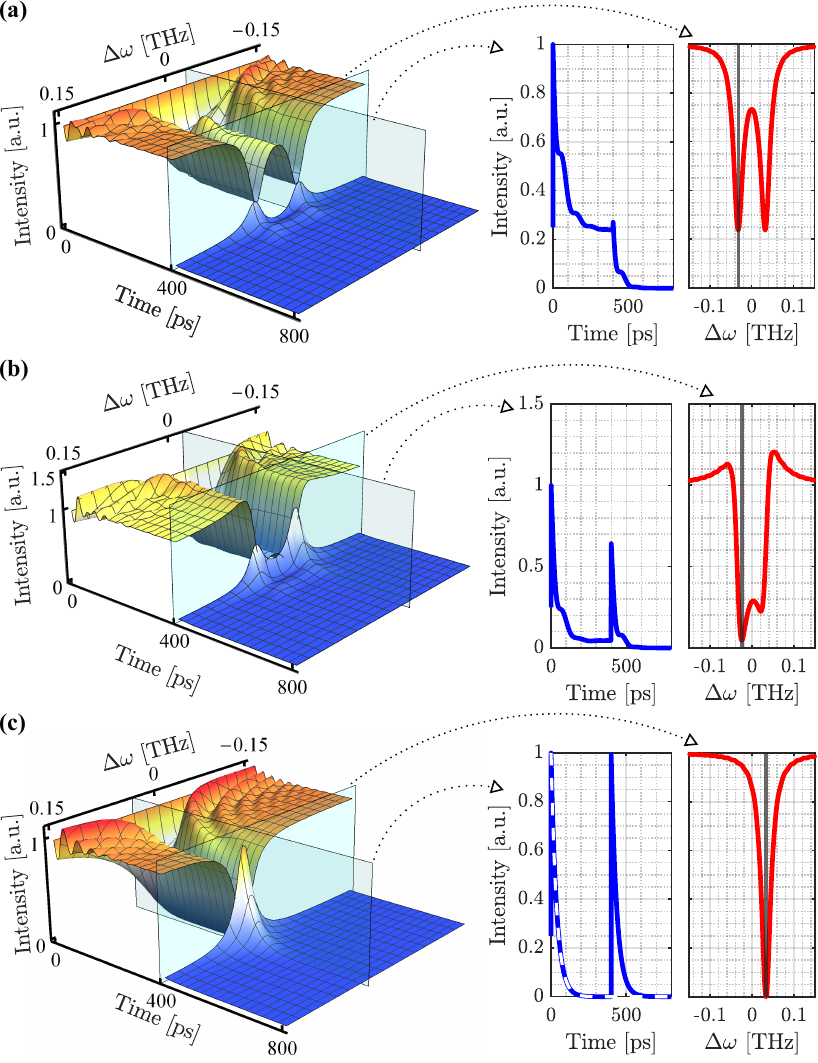}
	\caption{Intensity of the outgoing field from the right side for a Hermitian coupling. The first column shows the intensity as a function of time and frequency detuning ($\Delta \omega$). The second column shows the spectral and temporal line scan at the transparent planes in the 3D graphs. Panel (a) reports a single side excitation, while panels (b) and (c) show a symmetric interferometric excitation. Here, we use the following coefficients: $\epsilon=1$ a.u., $\Gamma=\gamma=6.8$ GHz, $\beta_{12}=-\beta_{21}^*= \beta = 33.2$ GHz and $\Delta t=400$ ps. In panel (b) and (c) we set $\phi=\frac{8}{5} \pi$ and $\phi=-\frac{\pi}{2}-arg[\beta]$, respectively.}
	\label{fig:Hermitian}
\end{figure}

\begin{figure}[t!]
	\centering
	\includegraphics[scale=1]{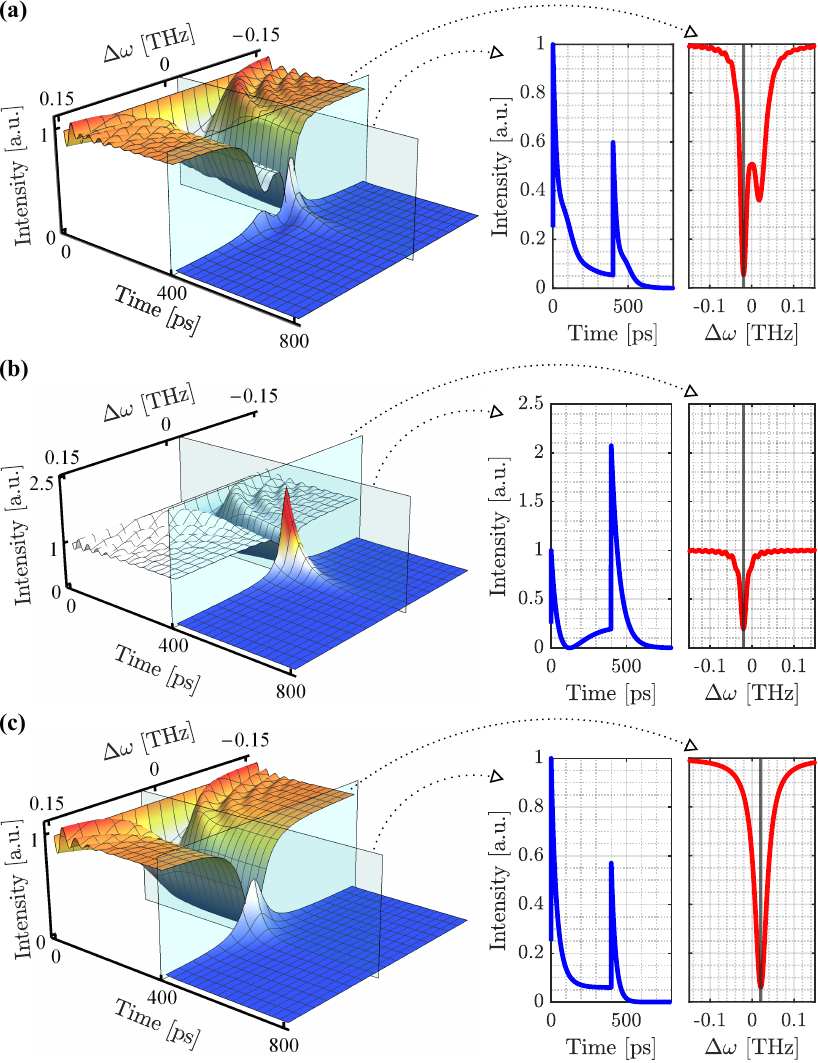}
	\caption{Intensity of $E_{\rm out,R}$ as a function of the frequency detuning $\Delta \omega$ and time for a non-Hermitian coupling. The first column shows $|E_{\rm out,R}|^2$ as a function of $\Delta \omega$ and time, while the second shows the spectral and temporal line scans at the transparent planes in the 3D graphs. The parameters used are: $\Gamma=\gamma=6.8$ GHz, $\beta_{12}=20.2$ GHz, $\beta_{21}=(-20.2+9\,i)$ GHz and $\Delta t=400$ ps. A single side excitation is shown in panel (a). Differently, in panels (b) and (c) is reported an interferometric excitation at the conditions where the doublet is merged into single Lorentzians.}
	\label{fig:Non-Hermitian}
\end{figure}

When the interferometric excitation is performed with square pulses of duration $\Delta t$ with angular frequency $\omega$ ($\xi_{\rm in}[t]=(\Theta[t]-\Theta[t-\Delta t])e^{- i \omega t}$), we explicitly get:
\begin{equation}
\begin{array}{r@{\ }c@{\ }l}
     \!\!E_{\rm out,R}[t]\!\!\!&=& \!\! e^{-i\omega t}\Big\{ \Theta[t] \Big\{ \epsilon _{\rm in,L}\!\! \left(\!\!1 + \frac{\Gamma (e^{-i t \sigma^{-}}\!\!\!-1)}{i \sigma^{-}} + \frac{\Gamma (e^{-i t \sigma^{+}}\!\!\!-1)}{i \sigma^{+}} \!\!\right) \\
     \!\!&+&\!\! \epsilon_{\rm in,R} \sqrt{\frac{\beta_{12}}{\beta_{21}}}e^{i \phi} \Gamma \!\left( \frac{e^{-i t \sigma^{+}}\!\!\!\!-1}{i \sigma^{+}} - \frac{e^{-i t \sigma^{-}}\!\!\!\!-1}{i \sigma^{-}}\right)\Big\} \\
     \!\!&-&\!\! \Theta[t \!-\! \Delta t] \! \Big\{\!\epsilon _{\text{in,L}} \!\! \left(\!\! 1 \!+\! \frac{\Gamma (e^{i (\Delta t-t) \sigma^{-}}\!\!\!\!\!-1)}{i \sigma^{-}} \! +\! \frac{\Gamma (e^{i (\Delta t -t) \sigma^{+}}\!\!\!\!\!-1)}{i \sigma^{+}} \!\!\right)\! \!\! \\
     \!\!&+&\!\! \epsilon_{\rm in,R} \sqrt{\frac{\beta_{12}}{\beta_{21}}}e^{i \phi} \Gamma \! \left(\!\!\frac{(e^{i (\Delta t -t) \sigma^{+}}\!\!\!\!\!-1)}{i \sigma^{+}}- \frac{(e^{i (\Delta t -t) \sigma^{-}}\!\!\!\!\!-1)}{i \sigma^{-}}\!\!\right)\!\!\! \Big\}\!\!\!\Big\},
\end{array}
\label{eq:TimeEoutGeneral}
\end{equation}
where the short hand terms $\sigma^{+}=\lambda_{1}-\omega$ and $\sigma^{-}=\lambda_{2}-\omega$.
Figures \ref{fig:Hermitian} and \ref{fig:Non-Hermitian} show the outgoing field intensity, $|E_{\rm out,R}|^2$, as a function of time and frequency detuning for Hermitian and non-Hermitian coupling, respectively. The first column shows the three-dimensional (3D) plots of $|E_{\rm out,R}|^2$ while the second exemplary line scans for $|E_{\rm out,R}[\Delta \omega]|^2$ at $t=399$ ps (red curves), i.e., close to the end of the square pulse, and $|E_{\rm out,R}[t]|^2$ for $\Delta \omega$ indicated by the vertical black line in the spectra (blu curves). In both \Figures{} \ref{fig:Hermitian} and \ref{fig:Non-Hermitian}, panel (a) shows the transmitted intensity for a single side excitation ($ \epsilon_{\rm in,L} = 1 $ a.u. and $ \epsilon_{\rm in,R} = 0 $), while panels (b) and (c) report an interferometric excitation ($ \epsilon_{\rm in,L} \neq 0 $ and $ \epsilon_{\rm in,R} \neq 0 $). 

\textit{The Hermitian coupling} occurs when $\beta_{12}=-\beta_{21}^*=\beta$. This condition ensures that the same energy is exchanged between the \clockwise{} and \counterclockwise{} modes. Consequently, the transmission spectrum for a single side excitation exhibits a balanced doublet. Here, with balanced doublet we mean a split resonance with the same extinction ratio for the two transmission dips \cite{Non-Hermitian_Biasi}, as shown in \Figure{} \ref{fig:Hermitian} (a). The coupling between the counter-propagating modes induces a time evolution toward the steady state characterized by intensity oscillations. Oscillations are present in the charging of the \microresonator{} as well as in its discharge (see temporal profile). As a result, the exchange of energy between the modes hides the exponential decays of an ideal resonator \cite{TimeResonator_Biasi}. Therefore, it makes erroneous an estimate of the \microresonator{} \qualityfactor{} by the \CavityRingDown{} method.

A symmetric interferometric excitation, i.e., $\epsilon _{\text{in,R}} = \epsilon _{\text{in,L}}= \epsilon$, allows changing the spectral shape of the doublet by varying the relative phase of the input fields \cite{Interferometric_BiaFran}. Here, it is useful to introduce the concept of super-modes to discuss the behavior of the two resonance doublet dips. As definition of them we consider the picture introduced in \cite{Interferometric_BiaFran}: $b_{1,2} = (\alpha_{\rm CCW} \pm \alpha_{\rm CW} e^{-i\phi})/\sqrt{2}$. Furthermore, we assume as a positive (negative) super-mode the one which exhibits a dip on a positive (negative) resonant frequency. As shown in \Figure{} \ref{fig:Hermitian} (b), by fixing $\phi = \frac{8}{5} \pi$ and with $\Gamma=\gamma$, a preferential exchange of energy towards a given super-mode is induced. 
This value of $\phi$ leads to a strongly unbalanced doublet (different spectral dips) and does not avoid the intensity oscillation during the time evolution (see the temporal profile). The asymmetric exchange of energy between the super-modes is shown in the discharge ($t>\Delta t$) as well. Indeed, when the inputs are turned off, the outgoing field exhibits peaks which are larger for the negative super-mode ($\Delta \omega < 0$) than for the positive one ($\Delta \omega > 0$) and which decay with different line-shapes. 

Panel (c) of \Figure{} \ref{fig:Hermitian} shows the outgoing field intensity for a symmetric interferometric excitation satisfying the condition where the spectral doublet merges into a perfect Lorentzian centered at $\Delta \omega =|\beta|$ \cite{Interferometric_BiaFran}. In this case, $\left|E_{\rm out,R}[t]\right|^2$ takes the form reported in \cite{TimeResonator_Biasi} for a single-mode \microresonator{}. In fact, using the conditions: $\epsilon _{\rm in,R} = \epsilon _{\rm in,L}= \epsilon$ and $\phi= \pm \frac{\pi}{2} - arg[\beta] + 2\pi m$ ($m\in \mathbb{Z}$), \equationM{} \eqref{eq:TimeEoutGeneral} reduces to: 
\begin{equation}
\begin{array}{r@{\ }c@{\ }l}
     \left|E_{\rm out,R}[t]\right|^2\!\!&=& \! \epsilon^2 \Big| \Theta[t] \frac{\gamma - \Gamma - i (\Delta \omega \pm |\beta|)+2 \Gamma e^{-t (-i(\Delta \omega \pm |\beta|)+ \gamma + \Gamma)}}{\gamma + \Gamma - i (\Delta \omega \pm |\beta|)} \\ &-&\!  \Theta[t- \Delta t]\!\\
     &\cdot&
     \frac{\gamma-\Gamma - i(\Delta \omega \pm |\beta|)+2\Gamma e^{-(t-\Delta t) (-i(\Delta \omega \pm |\beta|)+ \gamma +\Gamma)}}{\gamma +\Gamma - i (\Delta \omega \pm |\beta|)}\Big|^2\!\!.
\end{array}
\label{eq:TimeIntensity}
\end{equation}
Considering the discharge ($t>\Delta t$) of the \microresonator{},  \eqref{eq:TimeIntensity} reduces to:
\begin{equation}
\begin{array}{r@{\ }c@{\ }l}
     \left|E_{\rm out,R}[t]\right|^2\!\!&=& \! e^{-2(\gamma +\Gamma) t }
     \\ &\cdot&
     \frac{4 \epsilon^2 \Gamma^2 
     \left(1 + e^{2\Delta t(\gamma+\Gamma)}-2e^{\Delta t(\gamma+\Gamma)}\cos{\left[\Delta\omega \pm|\beta|\right]} \right)}
     {(\gamma+\Gamma)^2 + (\Delta \omega \pm |\beta|)^2}
     ,
\end{array}
\label{eq:Intensityt>Dt}
\end{equation}

namely, the product of a constant by an exponential decay function with a time constant $\tau_{ph} = \frac{1}{2 \left(\gamma + \Gamma \right)}$. Therefore, when the doublet merges into a single Lorentzian, there are no more oscillations due to the backscattering coupling of the counter-propagating modes. Therefore, a \CavityRingDown{} measurement allows estimating the \microresonator{} \qualityfactor{} due to only intrinsic and extrinsic factors. Interestingly, the discharge of the \microresonator{} follows a simple exponential decay also for $\Delta \omega \neq \mp |\beta|$. This result is quite appealing from an experimental point of view because it relaxes the requirement for high-precision spectral control of the input frequency in \CavityRingDown{}  measurements.
In the critical coupling regime, i.e., $\gamma = \Gamma =G$, imposing $\Delta \omega = \mp |\beta|$ \equationM{} \eqref{eq:TimeIntensity} reduces to an exponential law also during the \microresonator{} charging ($0\leq t\leq\Delta t$):
\begin{equation}
\left|E_{\rm out,R}[t]\right|^2 = \epsilon^2 e^{-4 G t}.
\label{eq:TimeExp}
\end{equation}
This behavior is represented by the dashed white line in the temporal profile of panel (c). Note that in the Hermitian regime, the two super-modes at $\Delta \omega = \mp |\beta|$ have the same exponential decay, and, therefore, an equal field enhancement or energy storage.

\textit{The non-Hermitian coupling} has no a-priori relation between $\beta_{12}$ and $\beta_{21}$. Here, we restrict to the case of a lossy \microresonator{}, e.g. silicon based ones.  
The counter-propagating modes exhibit an asymmetric exchange of energy \cite{Non-Hermitian_Biasi}. As shown in panel (a) of \Figure{} \ref{fig:Non-Hermitian}, this induces a strongly unbalanced transmission doublet in the steady state. Also, the temporal behavior shows intensity oscillations superimposed to an exponential decay (see the temporal profile in panel(a), right). Note that the non-Hermitian condition generates more stored energy into the negative super-mode with respect to the positive one. Consequently, when the source is switched off, the \microresonator{} discharges with an oscillating exponential decay. The temporal profiles do not allow an accurate \qualityfactor{} estimation in a common \CavityRingDown{} experiment.

An asymmetric interferometric excitation satisfying the condition: $\epsilon_{\rm in,R}=\sqrt{\frac{}{}|\beta_{21}|/|\beta_{12}|} \epsilon_{\rm in,L}$ and $\phi = (\pi/2 \pm \pi/2) + (arg[\beta_{21}]-arg[\beta_{12}])/2+2\pi m$  ($m\in \mathbb{Z}$), makes the doublet to merge into a single Lorentzian \cite{Interferometric_BiaFran}. This is shown in panels (b) and (c) of \Figure{} \ref{fig:Non-Hermitian}. The temporal decay lineshapes show no more the intensity oscillations at the resonant frequencies. Furthermore, for $\Delta \omega <0$ (panel (b)) the typical behavior of a single-mode \microresonator{} in the over-coupling regime is observed \cite{TimeResonator_Biasi}. In fact, the outgoing intensity goes to zero before the steady state and rises to a value larger than 1 in the discharge. In contrast, for $\Delta \omega >0$ (panel (c)) the temporal lineshape exhibits a pattern typical of a single-mode \microresonator{} in the under-coupling regime \cite{TimeResonator_Biasi}. The different coupling regimes are appreciated also in the discharging times. With the source off, the decay of the outgoing intensity follows an exponential law. 
As shown in  panels (b) and (c) of \Figure{} \ref{fig:Non-Hermitian}, the negative super-mode has more stored power than the positive one. Furthermore, they exhibit a different decay time constant, and thus, a different FWHM of their resonance spectra. Note that by changing the phase of $\beta_{12}$ and $\beta_{21}$ it is possible to swap between the under/over temporal evolution of the positive/negative super-modes \cite{Interferometric_BiaFran}.

To prove that the two super-modes assume the behavior of a single-mode \microresonator{} we write the eigenvalues of \eqref{eq:TCMT} as follows \cite{Interferometric_BiaFran}:
\begin{equation}
\begin{array}{r@{\ }c@{\ }l}
     \lambda_{1,2} &=& \omega_{0}\mp \tilde{\beta}- i\left(\Gamma+\gamma\mp \tilde{\gamma}\right),\\
        \tilde{\beta}&=&\sqrt{|\beta_{12}||\beta_{21}|}\,\sin{[(\varphi_{12}+\varphi_{21})/2]},\\
    \tilde{\gamma}&=&\sqrt{|\beta_{12}||\beta_{21}|}\,\cos{[(\varphi_{12}+\varphi_{21})/2]},
\end{array}
\label{eq:EigenvalueSinecosine}
\end{equation}
where $\varphi_{12/21}= arg[\beta_{12/21}]$, and thus, $\beta_{12/21}=\left|\beta_{12/21}\right|e^{i \varphi_{12/21}}$. Here,  $\tilde{\beta}$ and $\tilde{\gamma}$ are the real and imaginary parts of the backscattering coefficients. Precisely, the real part is the frequency shift while the imaginary one is another damping rate for the super-modes. Using \equationM{} \eqref{eq:EigenvalueSinecosine} and the condition for merging the doublet (i.e., $\phi = (\pi/2 \pm \pi/2) + (\varphi_{21}-\varphi_{12})/2)$) in \equationM{} \eqref{eq:TimeEoutGeneral} one obtains: 
\begin{equation}
\begin{array}{r@{\ }c@{\ }l}
     E_{\rm out,R}[t]\! &=& \!\frac{\epsilon_{\rm in,L} e^{-i \omega t}}{i (\gamma + \Gamma \mp \tilde{\gamma})+\Delta \omega \pm \tilde{\beta}} \Big\{\!\Theta[t] \!\Big(2 i \Gamma e^{it(\Delta \omega \pm \tilde{\beta}+ i (\gamma+ \Gamma \mp \tilde{\gamma}))}\\ &+&
     i(\gamma-\Gamma \mp \tilde{\gamma})+\Delta \omega \pm \tilde{\beta}\Big) \\ &-&
     \Theta[t-\Delta t] \Big( 2 i \Gamma e^{i(t-\Delta t)(\Delta \omega \pm \tilde{\beta}+ i (\gamma + \Gamma \mp \tilde{\gamma}))}\\ &+& i(\gamma - \Gamma \mp \tilde{\gamma}) + \Delta \omega \pm \tilde{\beta}\Big)\Big\}.
\end{array}
\label{eq:TimeEout_Val}
\end{equation}
This expression is analogous to the expression for the field transmitted by a bus waveguide coupled to a single-mode \microresonator{} unaffected by backscattering and excited from one side \cite{TimeResonator_Biasi}. Note that, in this formalism, the terms related to losses are expressed by means of the imaginary part. In the discharging ($t > \Delta t$), the field intensity given by \equationM{} \eqref{eq:TimeEout_Val} reduces to the product of a constant by an exponential function: 
\begin{equation}
\begin{array}{r@{\ }c@{\ }l}
     \left|E_{\rm out,R}[t]\right|^2\!\!\!&=& \! e^{-2(\gamma +\Gamma\mp\tilde{\gamma}) t }
     \\ &\cdot& 
     \frac{\!4 \epsilon_{\rm in,L}^2 \! \Gamma^2 
     \! \left(\!1 + e^{2\Delta t(\gamma+\Gamma\mp\tilde{\gamma})}-2e^{\Delta t(\gamma+\Gamma\mp\tilde{\gamma})}\!\cos{\left[\Delta\omega \pm\tilde{\beta}\right]} \! \right)}
     {(\gamma+\Gamma\mp\tilde{\gamma})^2 + (\Delta \omega \pm \tilde{\beta})^2}\!.
\end{array}
\label{eq:nonH-Intensityt>Dt}
\end{equation}
As for the Hermitian case, the exponential decay does not depends on $ \omega$. Note that in \eqref{eq:TimeEout_Val}, the intrinsic and extrinsic coefficients are related to $\tilde{\gamma}$. Specifically, $\tilde{\gamma}$ is subtracted in the case of the negative super-mode ($\Delta \omega = -\tilde{\beta}$) and added for the positive one ($\Delta \omega = \tilde{\beta}$). As a result, the field intensity of the two super-modes exhibits two different behaviors. In the critical coupling regime, at $\Delta \omega = \pm \tilde{\beta}$, the intensity given by \equationM{} \eqref{eq:TimeEout_Val} can be written as:
\begin{equation}
\begin{array}{r@{\ }c@{\ }l}
     |E_{\rm out,R}[t]|^2 &=& \left(\frac{\epsilon_{\rm in,L}}{(\pm \tilde{\gamma} - 2 G)}\right)^2 \Big\{ \Theta[t] \left(\pm \tilde{\gamma} -2 G e^{-t(2G\mp \tilde{\gamma})} \right) \\ &-&
     \Theta[t- \Delta t] \left( \pm \tilde{\gamma}-2G e^{-(t-\Delta t) (2 G \mp \tilde{\gamma})}\right)\Big\}^2  
\end{array}
\label{eq:TimeEoutCritical}
\end{equation}
By imposing $\tilde{\gamma} = 0$ this expression reduces to the Hermitian case of \equationM{} \eqref{eq:TimeExp}. As a result, the subtraction (sum) of $\tilde{\gamma}$ induces in the negative (positive) super-mode an under(over)-coupling regime. In \equationM{} \eqref{eq:TimeEoutCritical}, the exponential decay has $\tilde{\gamma}$ with a different sign in the over-coupling regime or in the under-coupling one, in accordance with the higher/lower stored optical energy. Along this discussion, the exponential \microresonator{} charging and discharging phases 
can be described in terms of the optical energy stored inside the \microresonator{} as $\zeta = \zeta[0] e^{-\frac{\omega}{Q} t}$, where $\zeta[0]$ is the optical energy stored at $t=0$. This can be related to the time dependence of the field intensity by considering \equationM{} \eqref{eq:TimeEout_Val}. From this comparison, one can extract \qualityfactor{} \cite{TaijiBack_Franchi}:
\begin{equation}
Q = \frac{\omega_0 \pm \tilde{\beta}}{2 (\gamma + \Gamma \mp \tilde{\gamma})}.
\label{eq:QualityFactor}
\end{equation}
Thus, the simple fit of the outgoing field intensity decay allows estimating the \qualityfactor{} of the system as in the Hermitian case. However, there is a difference between the estimated values obtained in the Hermitian and non-Hermitian coupling. In the former, 
$\tilde{\gamma} = 0$ and, consequently, \qualityfactor{} of both merged Lorentzians is independent of $\beta_{12/21}$. \qualityfactor{} is only related  to $\gamma$ and $\Gamma$ since $\tilde{\beta} \ll \omega_0$. 
As a result, we can estimate the ultimate \qualityfactor{}. In the non-Hermitian case, the different exchange of energy between the counter-propagating modes causes $\tilde{\gamma} \neq 0 $. The two super-modes exhibit different \qualityfactors{} and the backscattering term seems to induce even a more favorable condition by decreasing $\gamma$ (see \equationM{} \eqref{eq:QualityFactor}). This is a purely theoretical formal condition. In fact, a priori $\gamma$ is not isolated from $\tilde{\gamma}$.
However, it is always possible to get an underestimation of the ultimate \qualityfactor{} by considering the stronger of the two super-modes. Furthermore, the surface-wall roughness has a stochastic nature and, thus, it should generate a balanced/quasi-balanced transmission doublet \cite{Non-Hermitian_Biasi}.

\textit{In conclusion}, we propose a method to estimate the \qualityfactor{} of a high/ultra-high quality \microresonator{} in the presence of backscattering. This is based on the interferometric excitation measurement of the charging and discharging of the \microresonator{}, which we call interferometric \CavityRingDown{} technique. We show that by merging a typical transmission doublet into one of the two \microresonator{} super-modes yields a charging and discharging behavior typical of a single mode \microresonator{}. Consequently, the ultimate \qualityfactor{} of the \microresonator{} can be determined by an estimation of the photon lifetime. This result is demonstrated for conservative (Hermitian) and dissipative (non-Hermitian) coupling of the counter-propagating modes. Despite the discussion and modeling has been performed for an integrated ring resonator structure, the interferometric \CavityRingDown{} technique can be also generalized to other cavity geometries such as spherical \microresonators{}, bulk resonators or photonic crystals \microresonators{}.

\section*{Funding}
Ministero dell’Istruzione, dell’Università e della Ricerca [PRIN PELM (20177 PSCKT)].

\section*{Acknowledgments} 
S. Biasi acknowledges the co-financing of the European Union -  FSE-REACT-EU, PON Research and Innovation 2014-2020 DM1062 / 2021 and thanks Dr. Fernando Ramiro Manzano for the interesting discussions and especially the questions that arise from them. R. Franchi acknowledges the co-financing of PAT through the Q@TN joint lab.


\bibliographystyle{unsrt}
\bibliography{arXiv_Interferometric_CRD}

%

\end{document}